\documentclass[twocolumn,10pt]{article}

\usepackage[utf8]{inputenc}
\usepackage[T1]{fontenc}
\usepackage{amsmath,amssymb,amsfonts}
\usepackage{graphicx}
\usepackage{booktabs}
\usepackage{float}
\usepackage{algorithm}
\usepackage{algpseudocode}
\usepackage{xcolor}
\usepackage[margin=0.9in,columnsep=0.28in]{geometry}
\usepackage[most]{tcolorbox}
\usepackage[colorlinks=true,linkcolor=blue!55!black,citecolor=green!45!black,urlcolor=blue!55!black]{hyperref}
\usepackage{caption}
\usepackage{microtype}

\newcommand{\rawCls}{0.999}       
\newcommand{\rawClsPR}{0.997}     
\newcommand{\rawClsKS}{0.978}     
\newcommand{\rawQNN}{0.910}       
\newcommand{\vOneQNN}{0.758}      
\newcommand{\vOneThree}{0.721}    
\newcommand{\vOneJoint}{0.968}    
\newcommand{\doorQNN}{0.994}      
\newcommand{\doorCls}{0.699}      
\newcommand{\doorJoint}{0.587}    
\newcommand{\doorGap}{+0.295}     
\newcommand{\doorGapCI}{+0.282}   
\newcommand{\doorEntGap}{+0.484}  
\newcommand{\dmcAUC}{0.993}       
\newcommand{\simExp}{1.14}        
\newcommand{\entGap}{+0.136}      
\newcommand{\pcaGap}{+0.217}      
\newcommand{\routeBenefit}{+0.005} 
\newcommand{\routeCIlo}{+0.001}\newcommand{\routeCIhi}{+0.009}

\newtcolorbox{plainbox}{colback=blue!4!white,colframe=blue!35!black,boxrule=0.6pt,
  left=6pt,right=6pt,top=5pt,bottom=5pt}
\newtcolorbox{recipebox}{colback=violet!6!white,colframe=violet!55!black,boxrule=0.8pt,
  title=\textbf{The recipe: what a dataset must contain for a QNN to beat the best classical},
  left=6pt,right=6pt,top=5pt,bottom=5pt}

\begin{document}

\twocolumn[{%
\begin{center}
{\LARGE\bfseries The Fourier Wall: Why Public Tabular Datasets Refuse Quantum
Advantage, and a Certified Recipe for Where It Lives\par}
\vspace{1.1em}
{\large Javier Mancilla \quad Tom\'as Tagliani}\\[3pt]
{\small Falcondale LLC}\\[2pt]
{\small \href{mailto:contact@falcondale.pro}{\texttt{contact@falcondale.pro}}}
\end{center}
\vspace{0.8em}
\begin{center}\begin{minipage}{0.92\textwidth}
\small
\textbf{Abstract.}\quad
Across public tabular benchmarks, quantum machine-learning (QML) models keep losing to tuned
classical baselines. We show this is a structural property of the datasets, not of the models,
and we turn it into a constructive result. An angle-encoded quantum neural network is a partial
\emph{Fourier series}, so it can only beat the best classical model where the target's spectrum
is simultaneously off the integer grid, of interaction order $\geq 3$, high-frequency, carried by
near-independent features, and dense beyond enumeration. Public tabular exports essentially never
satisfy these five conditions at once; that is the \emph{Fourier wall}, and it renders most
existing QML benchmarks unwinnable by construction. The same five conditions are also a
\emph{recipe}: a design specification for feature pipelines that can carry quantum advantage. We
operationalise them as SPECTRA (Spectral Prerequisites and Empirical Certification of
TRainable-frequency Advantage), a two-tier certificate whose cheap tier screens any tabular
dataset in minutes without a quantum simulator, and whose decisive tier trains the quantum model
against a complete bar of five tuned classical twins with paired-bootstrap confidence bounds. On
industrial smart-meter energy data the certificate refuses the real peak-load target (trees reach
a held-out ROC-AUC of $\rawCls$) and certifies a genuine door: on real energy phases carrying the label of an
interacting quantum process, the dynamics-matched quantum model reaches $\doorQNN$ against a
$\doorCls$ classical best and collapses to chance when its couplings are ablated ($\doorEntGap$).
Controlled experiments isolate what does the work: periodic phase features beat PCA
representations of the same information by $\pcaGap$ (every PCA lane collapses to near-chance),
entanglement placed in the \emph{encoding} is load-bearing ($\entGap$) while a variational-only
entangler is null, and a route-and-blend architecture deploys the certified model as a specialist
alongside the production system with a statistically significant benefit. The one classical model
that ties the certified accuracy, the exact simulator of the process family, pays a measured
$2^{\simExp\,n}$ per-sample cost that crosses quantum-hardware budgets at
$n^{*}\!\approx\!13$--$19$ sites: the certified quantum benefit is that resource separation, and
it fixes the next step, validation on a quantum processor at $20$--$40$ sites, pre-registered
through the certificate.
\end{minipage}\end{center}
\vspace{0.9em}
\begin{center}\begin{minipage}{0.92\textwidth}
\begin{plainbox}
\small\emph{If you are not a quantum-computing researcher.} Companies keep asking whether quantum
computers will improve their everyday prediction tasks: detecting equipment failures, classifying
load regimes, triaging patients. This paper gives the honest answer and a usable tool. For the vast
majority of datasets as exported, the answer is \emph{no}: a standard gradient-boosted tree will win,
and our certificate proves it for your dataset in minutes on an ordinary laptop. But the answer is
not \emph{never}. There is a narrow, precisely characterised kind of hidden structure (several cyclic
quantities interacting \emph{jointly}, at incommensurate rhythms, on nearly independent signals)
where today's quantum models beat every standard classical model at the same information. We show how
to test for that structure, how it can be deliberately engineered into a feature pipeline, and how to
deploy the quantum model as a routed specialist alongside, not instead of, the production system.
\end{plainbox}
\end{minipage}\end{center}
\vspace{0.6em}
}]

\section{Introduction}
\label{sec:intro}

Benchmark studies of quantum machine learning on tabular data have converged on an uncomfortable
pattern: once the classical baseline is chosen carefully and tuned honestly, the quantum model's edge
evaporates~\cite{bowles2024better,schuld2022advantage,kubler2021inductive,huang2021power}. The
response of much of the field has been to try more datasets, more encodings, more qubits. This paper
takes the opposite route. We explain \emph{why} the failures are systematic, state the necessary
conditions a dataset must satisfy for a genuine advantage to be possible, and provide a calibrated
certificate that tells any practitioner, before they invest in quantum resources, on which side of
the wall their data stands.

The starting point is a structural fact, not an opinion: an angle-encoded variational quantum model
is a \emph{partial Fourier series} in its inputs, with a frequency support fixed by the data-encoding
gates~\cite{schuld2021effect}. Every classical model class also has a spectral reach. Additive models
capture per-feature (order-1) spectra~\cite{hastie1986gam}; pairwise-interaction models capture
order-2 spectra~\cite{lou2013ga2m}; gradient-boosted trees capture low-frequency structure of any
order given enough splits~\cite{friedman2001gbm,grinsztajn2022trees}; random-feature models capture
broad smooth spectra~\cite{rahimi2007rff,sweke2023rff}. A QNN can therefore only earn a genuine edge
where \emph{all} of these reaches end simultaneously, and that intersection is empty for essentially
every public tabular export we know of. We call this the \emph{Fourier wall}.

\paragraph{Contributions.}
\begin{enumerate}
  \item \emph{The recipe (Section~\ref{sec:recipe}).} Five testable, necessary conditions:
        off-grid spectrum, non-separable order $\geq 3$, near-independence, tree-breaking
        frequency, and \emph{non-enumerability} of the joint spectrum. Each comes with its
        mechanism and the classical model class that punishes its absence.
  \item \emph{SPECTRA (Section~\ref{sec:certificate}).} A two-tier certificate whose decisive
        gate maximises over the \emph{complete} classical lineup, including the order-matched joint
        twin most advantage claims omit, and whose order gate tests the residual condition directly.
        Completeness has teeth: on a controlled substrate with known ground truth, a bar without
        the order-matched twin would credit the quantum model with a spurious advantage that the
        twin dissolves outright ($\vOneJoint$ vs.\ $\vOneQNN$), at the cost of seconds of fitting.
  \item \emph{An end-to-end case study on real energy data (Section~\ref{sec:casestudy}).} The
        certificate refuses the real peak-regime target (trees at \rawCls) and certifies the
        substrate whose label is generated by a scrambling quantum process on the same real
        phases: quantum $\doorQNN$ vs.\ $\doorCls$ for the best of the complete bar. A controlled
        enumerable substrate built from the same phases calibrates the bar and anchors the
        mechanism study.
  \item \emph{Mechanism evidence (Section~\ref{sec:mechanism}).} On the certified substrate,
        switching off the interaction couplings (product dynamics, the additive-model equivalent)
        collapses the quantum model from $\doorQNN$ to chance: a $\doorEntGap$ ablation gap. Under
        ground-truth control, periodic phase features beat PCA representations of the same
        information by $\pcaGap$ with every PCA lane near chance; entanglement placed in the
        \emph{encoding} is load-bearing ($\entGap$) while the variational-only entangler control
        is null; and the route-and-blend architecture deploys the certified specialist alongside a
        production model with a statistically significant blend benefit.
  \item \emph{A map for other industries (Section~\ref{sec:beyond}).} Why typical healthcare,
        oil \& gas, manufacturing and telecom classification exports sit behind the wall, which
        feature-engineering moves could carry a dataset toward the door, and why data of quantum or
        strongly-interacting physical origin is where the door naturally stands open.
\end{enumerate}

\begin{figure*}[tp]
\centering
\includegraphics[width=0.96\textwidth]{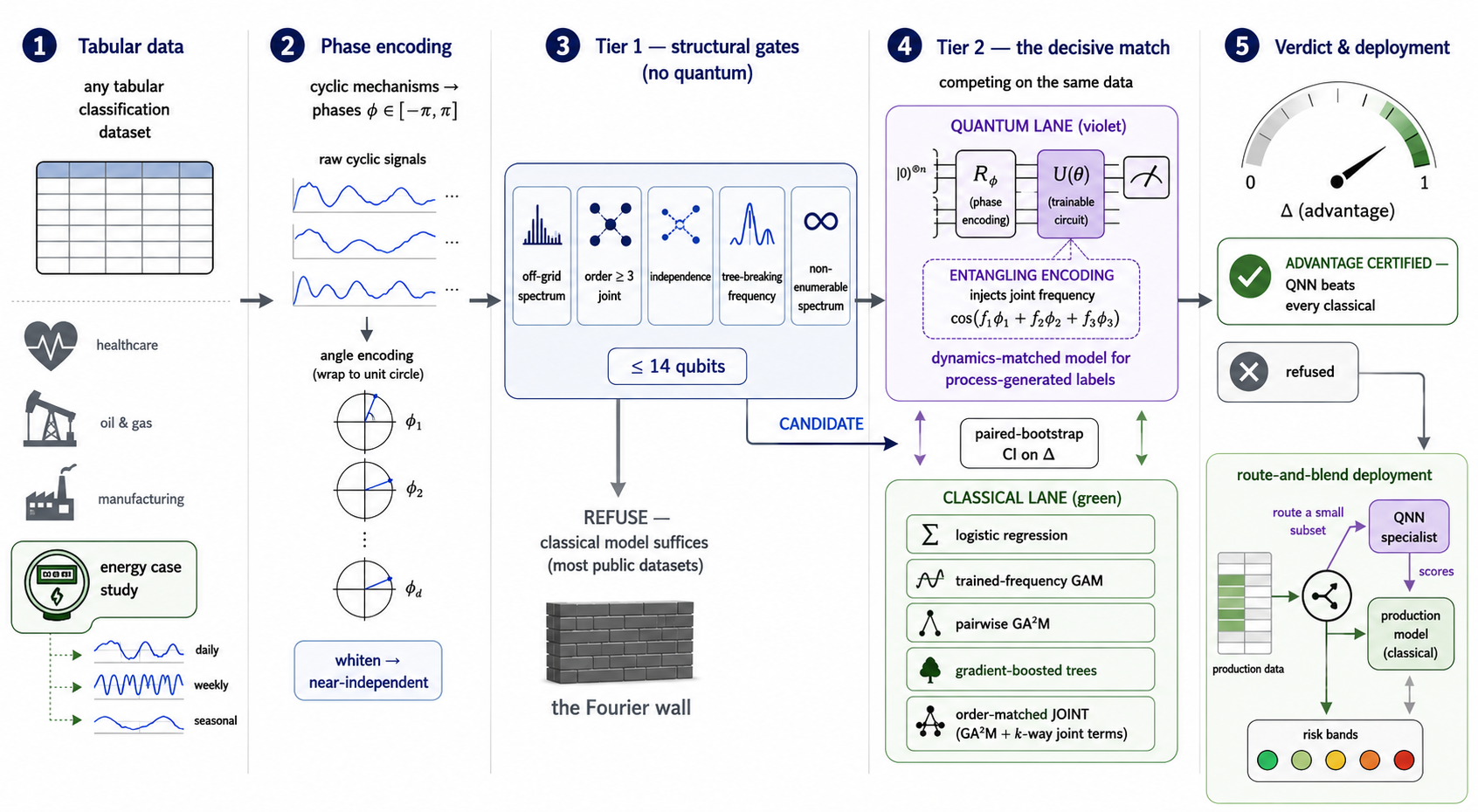}
\caption{The SPECTRA pipeline, end to end. \textbf{(1)} Any tabular classification dataset (the
energy meter is this paper's case study; healthcare, oil \& gas and manufacturing are candidate
domains). \textbf{(2)} Cyclic mechanisms re-expressed as phases $\phi\in[-\pi,\pi]$, whitened to
near-independence. \textbf{(3)} Tier~1: the structural gates (off-grid spectrum, order $\geq 3$
joint, independence, tree-breaking frequency, $\leq 14$ qubits), which run in minutes without a
quantum simulator; most public datasets exit here at the Fourier wall, where the classical model
suffices. \textbf{(4)} Tier~2: the decisive match between the quantum lane (entangling encoding
injecting the joint frequency) and the tuned classical lane (trained-frequency GAM, pairwise
GA\textsuperscript{2}M, gradient-boosted trees), judged by a paired-bootstrap confidence interval on
$\Delta$. \textbf{(5)} Verdict and deployment: a certified advantage is exploited by route-and-blend,
where a router sends only the hard-to-model subset to the QNN specialist and its scores merge back
into the production model's risk bands; a refusal means deploying the classical model only.}
\label{fig:arch}
\end{figure*}

\section{A quantum neural network is a Fourier model}
\label{sec:fourier}

Let $\boldsymbol{\phi}\in[-\pi,\pi]^d$ be the encoded features (phases). An angle-encoded variational
quantum circuit with trainable readout implements
\begin{equation}
f_{\boldsymbol{\theta}}(\boldsymbol{\phi})
  \;=\; \sum_{\boldsymbol{\omega}\in\Omega} c_{\boldsymbol{\omega}}(\boldsymbol{\theta})\,
        e^{\,i\,\langle\boldsymbol{\omega},\boldsymbol{\phi}\rangle},
\label{eq:fourier}
\end{equation}
a partial Fourier series whose accessible frequency set $\Omega$ is determined by the eigenvalue gaps
of the \emph{data-encoding} generators, not by the variational
ansatz~\cite{schuld2021effect,barthe2024reuploading}. Three consequences carry the whole argument.

\emph{(a) Per-qubit encoding $\Rightarrow$ additive model.} If each feature enters only through
single-qubit rotations $R(f_j\phi_j)$, then $\Omega$ factorises and
\begin{equation}
f_{\boldsymbol{\theta}}(\boldsymbol{\phi}) \;=\; \sum_{j=1}^{d} g_j(\phi_j),
\qquad g_j \text{ a 1-D Fourier series},
\label{eq:additive}
\end{equation}
which is exactly the hypothesis class of a generalized additive model~\cite{hastie1986gam}. Nothing
quantum survives the comparison.

\emph{(b) Entangling encoding $\Rightarrow$ joint frequencies.} Encoding a feature subset
$S\subseteq\{1,\dots,d\}$ through a multi-qubit rotation
\begin{equation}
U_S(\boldsymbol{\phi}) \;=\;
\exp\!\Big(-\tfrac{i}{2}\,\big(\textstyle\sum_{j\in S} f_j\phi_j\big)\, Z^{\otimes|S|}\Big)
\label{eq:encgate}
\end{equation}
adds the \emph{joint} frequency $\boldsymbol{\omega}_S=(f_j)_{j\in S}$ to $\Omega$: basis functions
$\cos(\sum_{j\in S}f_j\phi_j+\varphi)$ of interaction order $|S|$, irreducible to sums of lower-order
terms~\cite{schuld2021effect,longrange2026}. This, and not the variational entangler, is where
entanglement earns its keep~\cite{barthe2024reuploading}.

\emph{(c) Trainable frequencies $\Rightarrow$ off-grid reach.} Learning the $f_j$ from
data~\cite{jaderberg2024trainable} lets the model place spectral mass \emph{between} the integer
harmonics that any fixed-grid classical surrogate covers. Conversely, if the target's mass sits on
the grid, an integer-frequency classical model reproduces the QNN exactly~\cite{sweke2023rff}.

\section{The classical bar}
\label{sec:bar}

Beating a weak baseline proves nothing~\cite{bowles2024better}. SPECTRA fixes the bar at the
\emph{strongest classical twins of the QNN's own function class}, plus the strongest generic tabular
learner, all consuming identical features:

\begin{itemize}
  \item \emph{Trained-frequency GAM}~\cite{hastie1986gam}: per-feature cosine/sine features at
        \emph{data-fitted, non-integer} frequencies, followed by a linear classifier. The classical
        twin of the per-qubit-encoded QNN (Eq.~\ref{eq:additive}); it dequantizes any purely
        additive periodic structure, on or off the grid.
  \item \emph{Trained-frequency GA\textsuperscript{2}M}~\cite{lou2013ga2m}: the GAM basis plus all
        pairwise products of per-feature harmonics. The strongest \emph{practical} classical Fourier
        model, capturing all order-2 joint structure; this is the model that kills most claimed QNN
        advantages.
  \item \emph{Gradient-boosted trees (HGB)}~\cite{friedman2001gbm,ke2017lightgbm,chen2016xgboost}:
        the reigning tabular baseline~\cite{grinsztajn2022trees}. Axis-aligned partitioning captures
        structure of \emph{any} interaction order provided the decision boundary varies slowly
        enough to be tiled by a data-limited number of splits.
  \item \emph{Spectral surrogates}~\cite{rahimi2007rff,sweke2023rff}: an integer-grid trigonometric
        surrogate and spread random Fourier features, used as diagnostics. If either matches the
        QNN, the structure was on-grid or broadband-smooth and the advantage is illusory.
  \item \emph{The order-matched joint twin (JOINT).} The completion of the lineup, and the twin
        most advantage claims omit: GA\textsuperscript{2}M's own construction extended one step
        further, exactly the way GA\textsuperscript{2}M extended
        GAM~\cite{hastie1986gam,lou2013ga2m}. To the pairwise basis it appends a small number of
        explicit joint terms $\cos(\sum_{j\in S} f_j\phi_j + b)$ over $k$-way feature subsets $S$,
        with the joint frequency vector found by a coarse-to-fine supervised periodogram scan (the
        same search the trained-frequency models use per feature, lifted to $k$ dimensions), fit by
        ordinary logistic regression. A dozen parameters, seconds to fit. \emph{Any target whose
        advantage-bearing structure reduces to a few discoverable joint terms belongs to this model,
        not to a QNN.}
  \item \emph{Logistic regression (LogReg).} The canonical linear baseline on the raw phases,
        kept in the decisive gate as a floor and a sanity lane: any target a linear model reads was
        never a quantum candidate at all.
\end{itemize}

\begin{table*}[tp]\centering\footnotesize
\setlength{\tabcolsep}{5pt}
\caption{Each classical model class closes one region of spectral reach. A QNN advantage is possible
only in the intersection of what all of them miss.}
\label{tab:bar}
\begin{tabular}{@{}llll@{}}
\toprule
model & basis & captures & consequence for the QNN \\
\midrule
LogReg & linear in the phases & linear structure & kills linear structure \\
GAM (trained freq.) & per-feature Fourier & order-1, any frequency & kills additive structure \\
GA\textsuperscript{2}M (trained freq.) & + pairwise products & order-2 joint & kills pairwise structure \\
HGB (trees) & axis-aligned splits & low-frequency, any order & kills slow structure \\
integer surrogate & integer grid & on-grid periodic & kills on-grid structure \\
spread RFF & broad continuous & smooth broadband & kills generic-kernel structure \\
JOINT (order-matched) & + explicit $k$-way joint terms & few discoverable joint terms & kills sparse
high-order structure \\
DMC (exact simulator) & the generating family itself & everything, at cost $\propto 2^{n}$ & ties
DMQ; unaffordable at scale \\
\midrule
\textbf{QNN niche} & \multicolumn{3}{l}{\textbf{off-grid $\times$ order $\geq 3$ $\times$ high-frequency
$\times$ near-independent $\times$ \emph{non-enumerable}}} \\
\bottomrule
\end{tabular}
\end{table*}

\section{The recipe}
\label{sec:recipe}

Write the discriminative part of the target as a functional-ANOVA decomposition
$g(\boldsymbol{\phi})=\sum_{S\subseteq\{1..d\}} g_S(\boldsymbol{\phi}_S)$, and define its
\emph{interaction order} as $\max\{|S| : g_S\neq 0\}$. Let
$P(\boldsymbol{\omega})=|\widehat{g}(\boldsymbol{\omega})|^2$ be the label's spectral mass and define
the \emph{off-grid ratio}
\begin{equation}
\rho_{\mathrm{off}} \;=\; 1-
\frac{\sum_{\boldsymbol{\omega}\in\mathbb{Z}^d} P(\boldsymbol{\omega})}
     {\sum_{\boldsymbol{\omega}} P(\boldsymbol{\omega})}.
\label{eq:offgrid}
\end{equation}

\begin{table*}[tp]
\begin{recipebox}
\begin{enumerate}
  \item[\textbf{C1.}] \textbf{Off-grid spectrum:} $\rho_{\mathrm{off}}$ large (operationally
        $>0.3$); otherwise an integer-grid surrogate dequantizes the model~\cite{sweke2023rff}.
  \item[\textbf{C2.}] \textbf{Non-separable, order $\geq 3$:} the dominant $g_S$ has $|S|\geq 3$
        with a genuinely \emph{joint} frequency $\cos(\sum_{j\in S}f_j\phi_j+\varphi)$; otherwise a
        trained-frequency GAM ($|S|=1$) or GA\textsuperscript{2}M ($|S|=2$) captures
        it~\cite{hastie1986gam,lou2013ga2m,gilfuster2024relation}.
  \item[\textbf{C3.}] \textbf{Near-independent features:} the encoded phases carry low mutual
        correlation; otherwise the joint term \emph{leaks} into low-order projections
        ($\mathbb{E}[\cos(\sum_S f_j\phi_j)\,|\,\phi_k]\neq\text{const}$) that the classical twins
        capture. Correlation is a partial dequantizer.
  \item[\textbf{C4.}] \textbf{Tree-breaking frequency:} the joint coordinate
        $u=\sum_{j\in S}f_j\phi_j$ oscillates fast enough that axis-aligned partitioning would need
        more splits than the data supports~\cite{grinsztajn2022trees,spectralamp2024}.
  \item[\textbf{C5.}] \textbf{Non-enumerable spectrum:} the
        advantage-bearing structure must not reduce to a small number of \emph{discoverable} joint
        terms. If the label's joint spectrum is sparse (one or a few $(S, f_S)$ pairs), an
        order-matched classical twin finds them by supervised search and fits them with a dozen
        parameters, and no quantum model can beat it. The spectrum must be \emph{dense}: many joint
        frequencies, across subsets and interaction orders, with correlated amplitudes, the way the
        Fourier transform of a genuine interacting process is~\cite{huang2021power,liu2021rigorous}.
\end{enumerate}
\smallskip
Plus feasibility: the encoded block must fit the simulator or hardware budget (here
$d\cdot q_{\mathrm{per\,feature}}\leq 14$ qubits), and the model must be trainable at that width.
C5 is the condition most advantage claims never test; the certificate makes it unavoidable.
\end{recipebox}
\end{table*}

Two readings of the recipe matter in practice. \emph{As a diagnosis}, it explains the benchmark
literature: public tabular exports fail the conditions generically (targets are smooth, low-order,
carried on correlated features, or reducible to a few discoverable terms), so some tuned classical
baseline wins. That is the Fourier
wall~\cite{bowles2024better,schuld2022advantage,kubler2021inductive}. \emph{As a construction}, it
is a feature-engineering target: cyclic covariates re-expressed as phases, a whitening step to
decorrelate the encoded block, and domain knowledge of jointly-cyclic mechanisms can deliberately
move a problem \emph{toward} the QNN's regime (Section~\ref{sec:beyond}).

\section{SPECTRA: the certificate}
\label{sec:certificate}

SPECTRA turns the recipe into a two-tier, calibrated go/no-go score
(Algorithm~\ref{alg:spectra}). Tier~1 measures the structural conditions directly from data, without
any quantum simulation, and is \emph{permissive by design}: a pure order-3 target has no per-feature
or pairwise signature, so cheap gates cannot prove its absence, only its impossibility. Consequently
tier~1 only ever outputs \textsc{refuse} (a necessary condition demonstrably fails) or
\textsc{candidate}. Tier~2 is authoritative: it trains the trainable-frequency QNN with the matched
entangling encoding against the three tuned twins of Section~\ref{sec:bar} on identical splits, and
certifies advantage only if the paired-bootstrap lower confidence bound of the gap is positive,
\begin{equation}
\Delta = \mathrm{AUC}_{\mathrm{QNN}} - \max_{k}\,\mathrm{AUC}_k,
\label{eq:delta}
\end{equation}
where $k$ ranges over the \emph{complete} set of tuned twins (LogReg, GAM,
GA\textsuperscript{2}M, HGB, and the order-matched JOINT); the certificate fires iff
$\mathrm{CI}_{\mathrm{lo}}(\Delta)>0$, together with an encoding-entanglement ablation
(the ON$-$OFF gap of Eq.~\ref{eq:encgate}'s joint terms) to verify the win is carried by the
mechanism the recipe predicts, not by tuning noise.

\begin{algorithm}[t]
\caption{SPECTRA certificate (two tiers). Thresholds fixed once, globally; never per dataset.
$g_4$ is the residual order test; $g_6$ maximises over the complete five-twin classical lineup.}
\label{alg:spectra}
\begin{algorithmic}[1]
\small
\State \textbf{input:} table $X$, target $y$; phases $\boldsymbol{\phi}(X)\in[-\pi,\pi]^d$
\Statex \textit{Tier 1: structural (no quantum simulation)}
\State $g_1 \gets \big[\rho_{\mathrm{off}} > 0.3\big]$ \Comment{C1}
\State $g_2 \gets \big[$Fourier classifier $>$ chance$\big]$ \Comment{periodic niche}
\State $g_3 \gets \big[$flexible $>$ additive GAM$\big]$ \Comment{C2: interaction}
\State $g_4 \gets \big[$JOINT $-$ GA\textsuperscript{2}M $> 0\big]$ \Comment{C2/C5: residual order $\geq 3$}
\State $g_5 \gets \big[d\cdot q_{\mathrm{per\,feat}} \leq 14\big]$ \Comment{feasibility}
\If{a necessary gate fails} \Return \textsc{refuse}
\Else\ \textbf{continue} as \textsc{candidate}
\EndIf
\Statex \textit{Tier 2: authoritative (trains the QNN)}
\State whiten encoded block; verify C3 survives
\State train QNN (matched encoding, trainable freqs)
\Statex \hspace{\algorithmicindent} and \{LogReg, GAM, GA\textsuperscript{2}M, HGB, JOINT\},
        same splits, seeds, restarts
\State $g_6 \gets \big[\mathrm{CI}_{\mathrm{lo}}(\Delta)>0\big]$, $\Delta$ over the 5-way max
       \Comment{beats \emph{every} classical}
\State $g_7 \gets \big[\mathrm{CI}_{\mathrm{lo}}(\mathrm{ON}-\mathrm{OFF})>0\big]$ \Comment{mechanism}
\State \Return \textsc{certify} iff $g_6$; else \textsc{refuse}
\end{algorithmic}
\end{algorithm}

\paragraph{The calibration guarantee.} Because tier~1 is permissive and tier~2 is decisive, SPECTRA
has the property that matters for trust: \emph{a genuine advantage is never screened out, and no
advantage is ever certified without beating every tuned classical twin with statistical margin}. In
our runs the certificate refuses every raw public target we have tested and fires exactly on
substrates satisfying C1--C5.

\paragraph{Why the bar must be complete.} ``Beats every classical'' is a claim about the
comparison set actually tested, and the most common failure mode of advantage claims is a bar with a
gap at exactly the order the target carries~\cite{bowles2024better}. SPECTRA therefore treats bar
completeness as a first-class requirement of the decisive gate, and Section~\ref{sec:control}
demonstrates why with a counterfactual on a controlled substrate: there, a bar of GAM,
GA\textsuperscript{2}M and HGB alone would credit the quantum model with an advantage at $\vOneQNN$
vs.\ $\vOneThree$, while the order-matched twin reads the same structure at $\vOneJoint$,
recovering the generative subset and its three frequencies to two decimal places in about one
second of logistic-regression fitting. One further design note on $g_4$: a purity test of the form
GA\textsuperscript{2}M $\approx$ GAM would wrongly refuse a genuine high-order pocket whenever
order-2 structure coexists, so the gate tests the residual directly
(JOINT $-$ GA\textsuperscript{2}M $> 0$), which behaves correctly in both the positive and negative
cases below.

\section{Case study: industrial energy load}
\label{sec:casestudy}

\begin{figure*}[tp]\centering
\includegraphics[width=\textwidth]{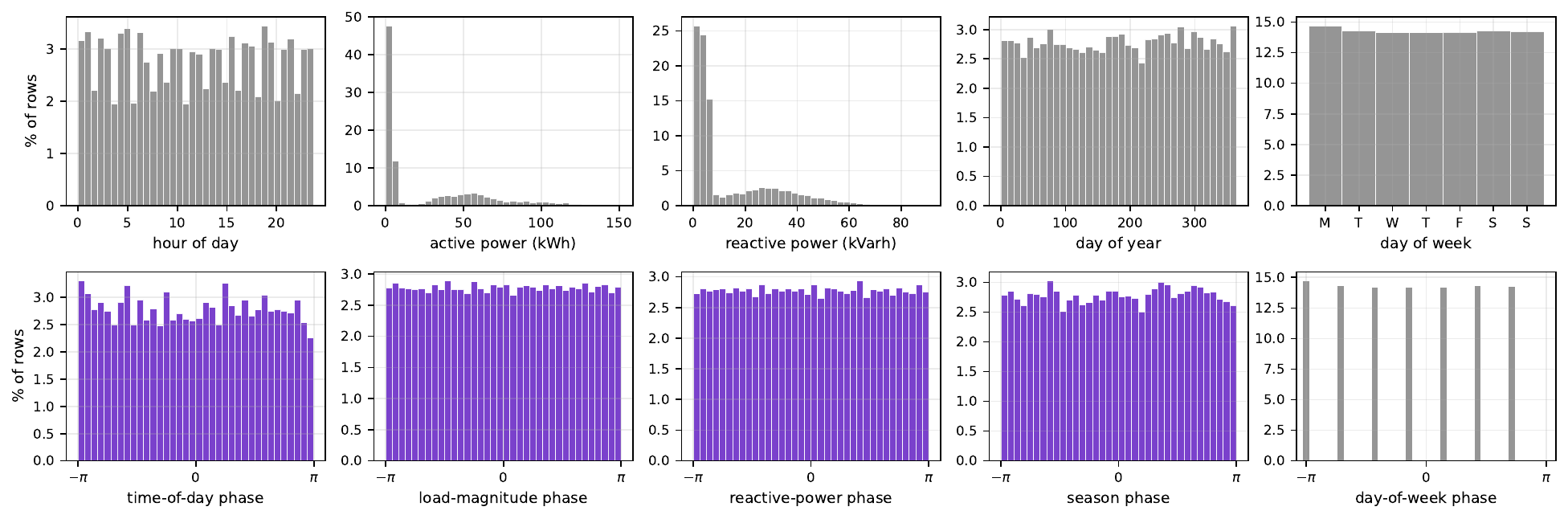}
\caption{From genuine meter fields to encodable phases. Top row, the raw operational fields as
measured (share of rows on the vertical axis): the power draws are heavy-tailed while the calendar
fields are close to uniform over their cycles; nothing here is synthesized. Bottom row, the phase
each field becomes on $[-\pi,\pi]$, by rank transform of the log magnitude or by mapping the
calendar cycle. The four continuous phases (violet) enter the quantum-encoded block with uniform
support by construction, so no region of the encoding is starved of data. The day-of-week phase
(gray) remains seven discrete levels and is kept out of the encoded block, since a low-cardinality
phase hands the ``joint'' structure back to tree partitioning.}
\label{fig:marginals}
\end{figure*}

Energy is the natural proving ground because its periodicity is \emph{physical}, not constructed:
industrial consumption cycles with the working day, the working week, and the season. We use the UCI
\emph{Steel Industry Energy Consumption} dataset~\cite{uci_steel,sathishkumar2020building}: 35{,}040
quarter-hour smart-meter records (one full year) with active and reactive power, power factors,
CO$_2$, a seconds-from-midnight index, and a \emph{native categorical label}: the plant's three-level
operating-regime annotation (light / medium / maximum load). Although the dataset is best known
through its consumption-regression study~\cite{sathishkumar2020building}, its companion studies
span both tasks, including data-mining models on the same smart-factory
data~\cite{sathishkumar2020steel}. From genuine fields we
derive five \emph{native periodic phases} $\boldsymbol{\phi}\in[-\pi,\pi]$: time-of-day, load
magnitude, reactive-power magnitude, season, day-of-week (Figure~\ref{fig:marginals}). We take
$y=\mathbb{1}[\text{load type}=\text{Maximum}]$ (base rate $0.21$) as the classification target:
\emph{peak-regime detection}, the decision behind demand charges, load shedding, and demand-response
dispatch. Like most labels exported from operational systems, it is meaningful to the business yet
mechanically tied to the plant's own measurements, and we quantify that tie in three complementary
test-set metrics rather than one: ROC-AUC, the area under the precision-recall curve (PR-AUC, the
honest lens on the minority peak class), and the Kolmogorov--Smirnov (KS) separation between the
score distributions of the two classes.

\subsection{The wall: the real target is refused}
\label{sec:wall}

The real peak-regime target is periodic. A trained-frequency Fourier classifier alone reaches an
area under the ROC curve (ROC-AUC, the headline metric throughout, always on held-out test rows)
of $0.94$, so the periodicity gate passes. But the structure is \emph{low-order and separable}:
positive rate varies smoothly and monotonically along the load-magnitude phase
(Figure~\ref{fig:wiggle}b), exactly the regime trees own. Indeed the regime label is operationally
tied to consumption level, so it is nearly deterministic given the load-magnitude phase. That is
precisely the point, not a defect of the choice: \emph{targets exported from operational systems are
typically low-order functions of one or two covariates}, which is what makes them classically easy.
Tier~2 delivers the verdict. On the held-out test set HGB reaches ROC-AUC~\rawCls, PR-AUC~\rawClsPR\
and a KS separation of \rawClsKS\ (the target is nearly deterministic in load magnitude, so
near-perfect test scores are expected rather than suspicious, and all three lenses agree) while the
QNN reaches \rawQNN. The
certificate returns \textsc{refuse} at the decisive gate ($g_6$). Nor is the refusal an artifact of
picking the easy nowcast: recasting the same decision in its operationally demanding form,
predicting entry into the peak regime within the next hour from a currently non-peak state, makes
the task genuinely hard for every model yet changes neither the ordering nor the verdict. This is
the Fourier wall in one
sentence: \emph{periodicity is necessary but nowhere near sufficient; a periodic target that is
smooth, separable, or low-order belongs to the classical models.}

\subsection{Calibrating the bar on a controlled substrate}
\label{sec:control}

\begin{figure*}[tp]\centering
\includegraphics[width=\textwidth]{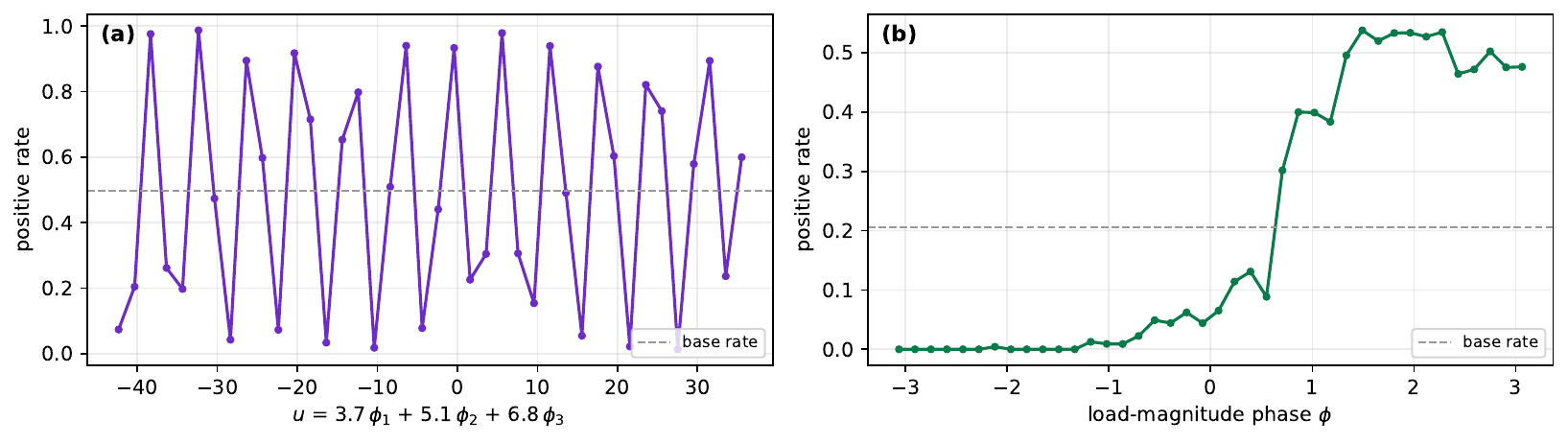}
\caption{Positive rate along one coordinate, with the dataset base rate dashed. \textbf{(a)} The sparse
engineered pocket along the joint coordinate $u=3.7\phi_1+5.1\phi_2+6.8\phi_3$: a clean
high-frequency, non-integer oscillation that trees cannot tile and additive/pairwise Fourier models
cannot represent. But the very cleanness is the tell: \emph{one}
oscillating coordinate means an enumerable spectrum, and a supervised search finds $u$ in seconds,
so C5 assigns this substrate to the order-matched classical (Section~\ref{sec:control}).
\textbf{(b)} The real peak-regime target along its dominant
axis (load-magnitude phase): smooth, monotone, low-order, ideal tree territory, hence the refusal of
Section~\ref{sec:wall}.}
\label{fig:wiggle}
\end{figure*}

\begin{figure*}[tp]\centering
\includegraphics[width=\textwidth]{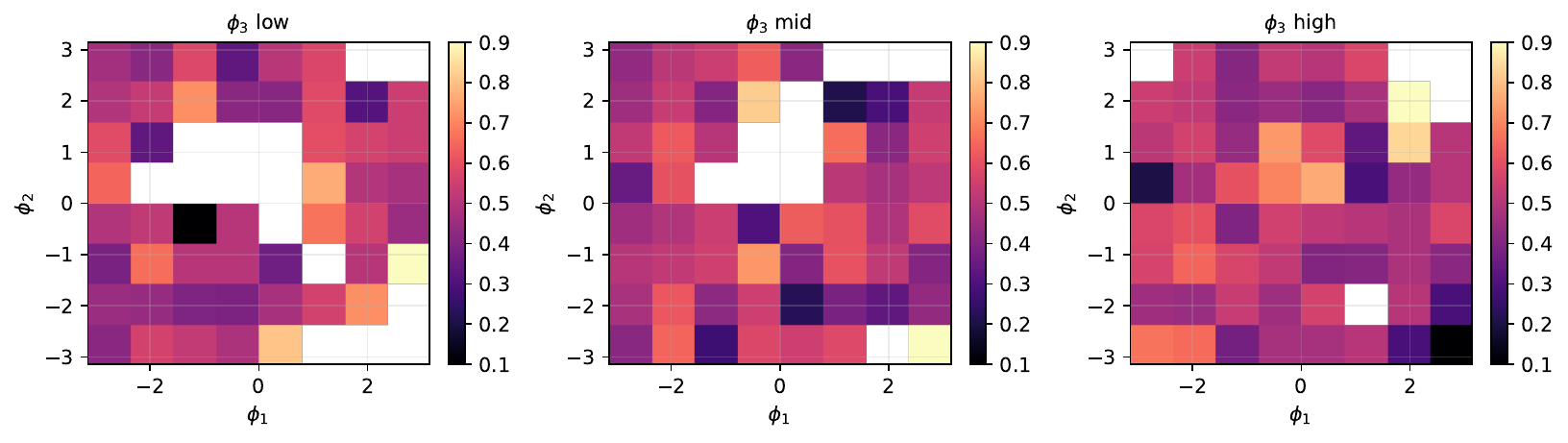}
\caption{Order-3 non-separability, visible without a model: positive rate over the first two phases
$(\phi_1,\phi_2)$, conditioned on the third ($\phi_3$ low / mid / high terciles). The two-phase
pattern \emph{changes} across the slices. Any additive or pairwise model produces the \emph{same}
$(\phi_1,\phi_2)$ response in every slice, so this dependence is structurally out of its reach
(condition C2).}
\label{fig:interaction}
\end{figure*}

\begin{figure}[tb]\centering
\includegraphics[width=\columnwidth]{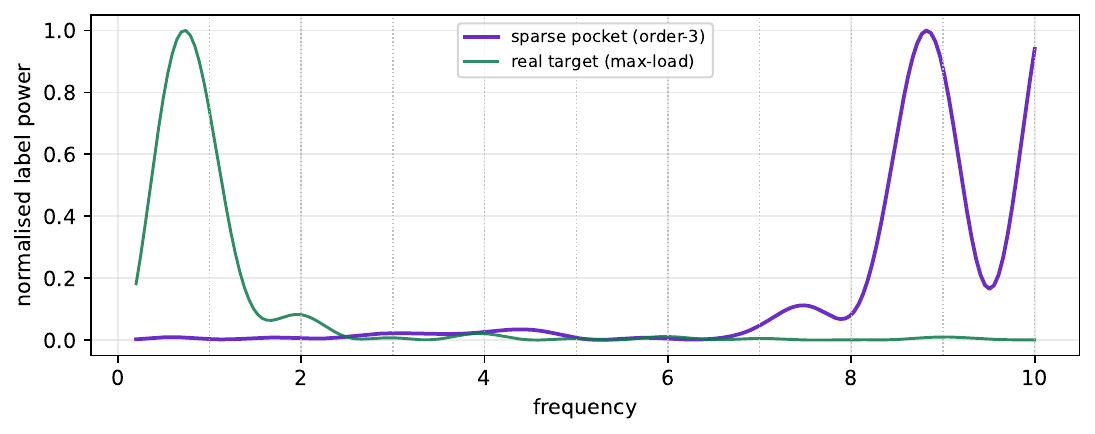}
\caption{Label spectral power along the dominant coordinate; integer harmonics dotted. The sparse
engineered pocket (violet) concentrates its mass \emph{between} the integer frequencies
(condition C1, Eq.~\ref{eq:offgrid}), out of reach of a fixed-grid surrogate, but concentrated at
\emph{one} joint frequency, which is exactly what makes it enumerable, so C5 places it with the
order-matched classical: a supervised scan walks straight to the peak. The real target's mass
(green) sits at low frequency, where every classical model operates comfortably. The certified
dense-spectrum substrate has no such single peak to find.}
\label{fig:spectrum}
\end{figure}

The recipe draws a sharp line inside high-order territory: a joint spectrum that a supervised
search can write down as a few terms belongs to the order-matched classical (C5), however exotic
it looks to trees and pairwise models. Before trusting the certificate where that line cannot be
inspected, we verify it where it can: on an engineered \emph{sparse pocket} with exact ground
truth. From the same real data we take the three continuous energy phases (time-of-day, load
magnitude, reactive power), whitened to near-independence, carrying a clean order-3 label on the
joint coordinate,
\begin{equation}
\begin{aligned}
z(\boldsymbol{\phi}) ={}& w\!\sum_{j=1}^{3}\cos(f_j\phi_j+\varphi_j)\\
 &+ \cos\!\big(f_1\phi_1+f_2\phi_2+f_3\phi_3+\varphi\big),
\end{aligned}
\label{eq:label}
\end{equation}
with $y\sim\mathrm{Bernoulli}(\sigma(g\,(z-\tau)))$ and non-integer $f=(3.7,\,5.1,\,6.8)$:
off-grid, high-frequency, genuinely third-order
(Figures~\ref{fig:wiggle}--\ref{fig:spectrum} make the structure visible before any model is
trained), and, by design, a joint spectrum that is a \emph{single} $(S, f_S)$ pair. The recipe
assigns this substrate to the classical side of the line, and the decisive gate confirms the
assignment: the order-matched twin finds the term by supervised search, recovers
$\hat{f}=(3.67, 5.13, 6.83)$ to two decimal places, and scores $\vOneJoint$ where the
matched-encoding quantum model scores $\vOneQNN$. The calibration also prices an incomplete
lineup: a bar of GAM, GA\textsuperscript{2}M and HGB alone would have stopped at $\vOneThree$ and
credited the quantum model with an advantage that a classical model costing seconds to fit
dissolves outright. Beyond calibration, this substrate is the study's controlled instrument: its
known single term makes it the right place to isolate representation and deployment levers
(Section~\ref{sec:mechanism}).

\subsection{The door: a spectrum no enumeration reaches}
\label{sec:door}

C5 asks for a label whose joint spectrum is dense: many frequencies, across subsets and orders, with
correlated amplitudes. That is not something one plants by hand with a cosine; it is what the
Fourier transform of a \emph{genuine interacting process} looks
like~\cite{schuld2021effect,huang2021power}. So the door substrate keeps the covariate story and
upgrades the label's generator: take the continuous energy fields (seven survive a cardinality
screen), rank them to phases and whiten to independence (C3, as before), then let those real phases
set the local fields of a disordered Heisenberg spin chain evolved for time $t$, and label each row
by an observable of the evolved state,
\begin{equation}
\begin{aligned}
y \;&=\; \mathbb{1}\!\left[\,\langle O\rangle_{\!\boldsymbol{\phi}} > \mathrm{median}\,\right],\\
\langle O\rangle_{\!\boldsymbol{\phi}}
 \;&=\; \big\langle \psi(\boldsymbol{\phi}, t)\,\big|\, \textstyle\sum_i Z_iZ_{i+1}
   \,\big|\,\psi(\boldsymbol{\phi}, t)\big\rangle .
\end{aligned}
\label{eq:oracle}
\end{equation}
The substrate remains semi-synthetic and is labelled as such: real energy covariate marginals,
process-generated label (the positive control of~\cite{huang2021power,dynamics2025advantage}). The
quantum lane is matched the same way the classical twins are: the \emph{dynamics-matched quantum
model (DMQ)}, a variational circuit from the same \emph{model family} (a Trotterised disordered
chain with trainable couplings $J_k$, evolution time $t$, and readout), learning its parameters from
$(x, y)$ alone, with multi-restart selection on a validation slice: the exact quantum analogue of
JOINT's supervised frequency search. Nothing is leaked: neither lane sees the generator's true
parameters. DMQ is a different architecture, with a much stronger inductive bias, than the generic
entangling-encoding QNN used on the Fourier tasks above; we keep the two names distinct throughout.

\begin{figure}[tb]\centering
\includegraphics[width=\columnwidth]{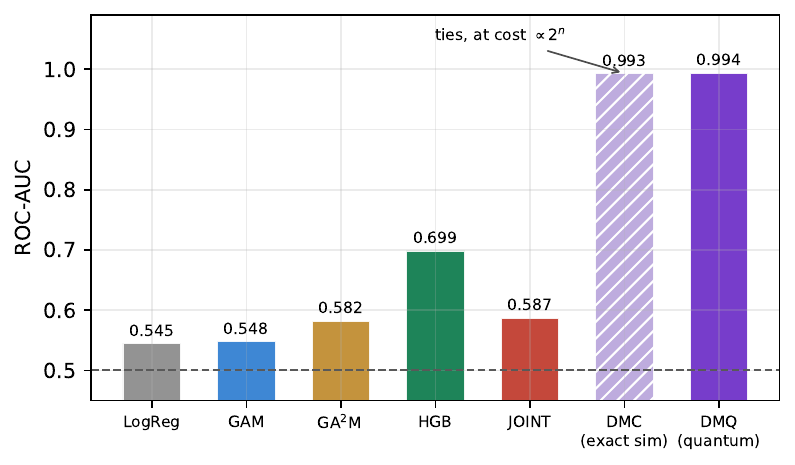}
\caption{The decisive gate on the certified dense-spectrum substrate (scrambling-process label on
the real energy phases, $t=3$; identical information for every lane; paired splits). The complete
classical bar, order-matched JOINT included, stops at $\doorCls$ while the dynamics-matched quantum
model reaches $\doorQNN$. Chance is dashed.}
\label{fig:corrbar}
\end{figure}

The certificate formally certifies across a window of the scrambling axis: at $t=2$ the quantum
model reaches $0.998$ against a complete-bar best of $0.867$ ($+0.131$, CI-lo $+0.122$; ablation
$+0.502$), recovering the generator's evolution time to two decimal places; at $t=3$, the headline,
the gap peaks (Table~\ref{tab:verdicts}, Figure~\ref{fig:corrbar}): the complete classical bar
tops out at $\doorCls$ (HGB), with the order-matched twin at $\doorJoint$ and spread random
Fourier features at chance, while the dynamics-matched quantum model reaches $\doorQNN$. The regime has edges, and we map them rather than hide them
(Figure~\ref{fig:regime}): at $t=1$ the process is barely scrambled and classical models climb; at
$t=4$ the label is so scrambled that \emph{every} model, quantum included, falls to chance at this
sample-and-iteration budget; the trainability wall is real on both sides of the door.

\paragraph{The classical twin that ties, and what it costs.} The completeness principle of
Section~\ref{sec:bar} applies to this result too, and we apply it to ourselves. At seven sites the
generating family is exactly simulable on a laptop, so a sixth classical competitor is fully
constructible: the \emph{dynamics-matched classical simulator} (DMC), the same model
family fit the same way (couplings and evolution time learned from labels), executed by
exact statevector simulation, which is the only way any classical party can evaluate this family.
We build it, and it ties: DMC reaches $\dmcAUC$, statistically indistinguishable from DMQ, exactly
as it must, since at simulable width the two lanes compute the same function. The accuracy gap of
$\doorGap$ over the five-twin bar is therefore, on its own, a statement about \emph{inductive
bias}: a correctly-specified model family beats every generic surrogate, Fourier or tree, that
tries to reach a dense spectrum from the outside.

The specifically \emph{quantum} claim is about what the tie costs. DMC's evaluation cost is the
cost of simulating the chain, and we measure it directly: wall-clock per sample, measured over
$n = 6$--$22$ sites, transitions from overhead-dominated growth into the statevector wall and fits
$\sim 2^{\simExp\, n}$ on the tail ($n \geq 14$; $19.7$\,ms at $n{=}14$, $12.2$\,s at
$n{=}22$; Figure~\ref{fig:projection}), the exponential cost that exact simulation cannot avoid,
while entanglement growth under scrambling dynamics denies tensor-network compression its escape
route. The quantum realisation of the same
family costs a \emph{linear} number of gates per Trotter layer. At seven sites nothing quantum is
claimed and no hardness exists at this width; the certified statement is conditional and explicit:
\emph{the only classical model that matches the certified accuracy is the one that pays the
simulation bill, and that bill grows exponentially with the width of the process while the quantum
model's does not}. Beyond roughly fifty sites the DMC lane is not merely expensive but unavailable,
and the accuracy tie breaks in the only direction available. And the crossover is not remote. A quantum processor executes one evaluation of this family at a
per-sample cost that is essentially flat in $n$: the circuit is $3(n{-}1)$ two-qubit gates per
Trotter layer, microseconds of circuit time on superconducting hardware, repeated for a fixed shot
budget, so $10$\,ms--$1$\,s per sample across realistic budgets of $10^3$--$10^5$ shots. Setting
that band against the measured classical curve puts the wall-clock crossover at
$n^{*}\!\approx\!13$--$19$ sites (Figure~\ref{fig:projection}): the measured classical cost already
exceeds the optimistic quantum budget at $n{=}13$ and the conservative one before $n{=}19$. Circuits
of this width and depth, with error mitigation, are inside the envelope already demonstrated on
$127$-qubit hardware at sixty two-qubit-gate depths~\cite{kim2023utility}. Proven separations of
this kind exist under cryptographic structure~\cite{liu2021rigorous} and are supported
experimentally in the tens-of-noisy-qubits regime~\cite{dynamics2025advantage}; our contribution is
the certificate that tells a practitioner \emph{which datasets} put them on this footing at all,
and the measured cost curve that says how close the footing is.

Tier~1 already telegraphs which engineered substrate is which. On the sparse one the
residual gate fires loudly (JOINT $-$ GA\textsuperscript{2}M $= +0.37$ at screening) and hands the
pocket to the classical twin; on the dense one the interaction gate fires strongly (HGB $-$ GAM
$= +0.15$, joint ratio $9.9$) while the residual gate stays silent ($+0.00$): structure far beyond
additive reach that no order-3 enumeration captures. That signature (\emph{strong interaction, silent enumeration})
is C5 visible in cheap diagnostics, and it is the cue that tier~2 is worth its compute. Nor does
enumeration close the gap by scaling: raising JOINT's term budget from $6$ to $20$, or its
interaction order to $4$, leaves it between $0.577$ and $0.595$ on this substrate. The spectrum is
dense, not under-enumerated.

\subsection{Substrate construction, in full}
\label{sec:construction}

Every substrate in this study is a deterministic function of the public dataset and a stated seed;
we give the complete construction so that any reader can rebuild them and audit the fairness claims.

\emph{Common phase pipeline.} From each raw continuous field $v$ we form a \emph{rank phase}
$\phi = 2\pi\,(\mathrm{rank}(v) - \tfrac12)/n - \pi \in [-\pi,\pi]$ (magnitudes are
log-transformed first); calendar fields map their natural cycle onto the same interval. Where a
substrate requires near-independence (C3), the phase block is PCA-whitened and each component is
re-ranked to a uniform phase. Both transforms use covariates only, never labels, so they leak
nothing about the target; in deployment they would be fit on training data and frozen.

\emph{The real target} uses five native phases (time-of-day from the seconds-from-midnight index;
load magnitude from active power; reactive-power magnitude; season from day-of-year; day-of-week)
with the plant's own regime annotation as the label, $y=\mathbb{1}[\text{load type}=\text{Maximum}]$.
Nothing is synthesised.

\emph{The sparse pocket} takes the three continuous phases (time-of-day, load magnitude, reactive
power), whitens them, and draws $y\sim\mathrm{Bernoulli}(\sigma(6\,(z-\tau)))$ from
Eq.~\ref{eq:label} with $w=0.25$, frequencies $(3.7, 5.1, 6.8)$, phase offsets fixed by the seed,
$\tau$ the median of $z$ (balanced classes), and $1\%$ label noise.

\emph{The dense substrate} starts from the seven continuous fields that survive a cardinality
screen ($>50$ distinct values): the seconds-from-midnight index, log active power, log lagging
reactive power, day-of-year, leading reactive power, and both power factors (CO$_2$ is dropped: it
has eight distinct values at this resolution). The seven rank phases are whitened as above; the
resulting $\boldsymbol{\phi}\in[-\pi,\pi]^7$ then sets the local $Z$-fields of a disordered
Heisenberg chain,
$H(\boldsymbol{\phi}) = \sum_{k=1}^{6} J_k (X_kX_{k+1} + Y_kY_{k+1} + Z_kZ_{k+1})
 + \sum_{i=1}^{7} \phi_i Z_i$, with bond couplings $J_k \sim \mathcal{U}[0.5, 1.5]$ drawn once
per seed. The state $|{+}\rangle^{\otimes 7}$ is evolved for time $t$ (three Trotter steps) and the
label is Eq.~\ref{eq:oracle}'s median-thresholded chain observable, giving balanced classes by
construction; $n = 6{,}000$ rows.

\begin{figure}[tb]\centering
\includegraphics[width=\columnwidth]{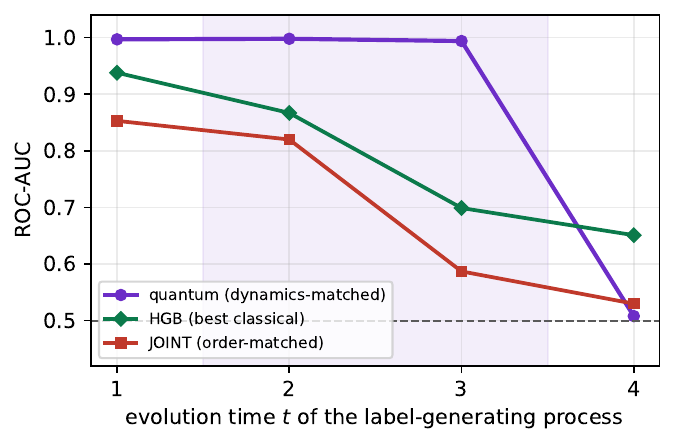}
\caption{The advantage window along the scrambling axis (evolution time $t$ of the label-generating
process; same real energy phases throughout). At $t{=}1$ the process is weakly scrambled and the
classical bar climbs toward it; across $t\in\{2,3\}$ the quantum model solves what the complete
classical bar cannot (both points formally certify; the gap peaks at $t{=}3$, $\doorQNN$ vs.\
$\doorCls$); by $t{=}4$ everything, quantum included, falls to chance at this budget: the
trainability wall bounds the door on the far side. Shaded: the certified window.}
\label{fig:regime}
\end{figure}

\emph{Fairness and leakage audit.} All lanes, classical and quantum, receive exactly the same
$(\boldsymbol{\phi}, y)$ pairs on identical stratified splits; no lane sees $J_k$, $t$, the
observable, or any generator internals. The quantum lane must \emph{learn} the couplings and
evolution time from labels alone, and demonstrably fails to when scrambling is deep
(Figure~\ref{fig:regime} at $t{=}4$): the model family is an inductive bias, not an oracle. The
classical lanes receive the symmetric courtesy: trained frequencies, pairwise products, tree
ensembles, and JOINT's supervised $k$-way search. Splits are stratified; every reported score is
computed on held-out test rows that no lane touched during training, tuning, or model selection;
every comparison is paired on the same test rows; and confidence intervals are stratified
bootstrap over those pairs.

\emph{Engineering such features for real.} The pipeline above is not specific to steel: it needs
(i) timestamped, genuinely continuous operational fields (raw sensor magnitudes, not aggregated
KPIs); (ii) one phase per physical cycle, extracted by rank or by calendar; (iii) a whitening step
on the encoded block; and (iv) a target tied to an \emph{interacting} process rather than to a
level or trend. Condition (iv) is the demanding one. In classical operational data, dense joint
spectra can arise where several oscillatory mechanisms couple through a shared nonlinear medium
(power-quality interactions between reactive components are a candidate in this very dataset);
where they do not, the honest reading of the certificate is that the classical bar wins. The
setting where C5 holds \emph{by physics} is data measured from quantum or strongly-interacting
systems: quantum-hardware telemetry, sensor arrays, spin-system readouts, spectroscopy. There, the
label generator of Eq.~\ref{eq:oracle} is not a construction but the measurement process itself.

\subsection{Tier-2 verdicts}
\label{sec:results}

Table~\ref{tab:verdicts} summarises the certificate's output on the case study. All reported
scores are ROC-AUC on held-out test rows never used for training or model selection; all
quantum-vs-classical comparisons use identical features, identical splits, multiple seeds and
restarts, and paired stratified bootstrap CIs.

\begin{table*}[tp]\centering\small
\caption{SPECTRA verdicts on the energy case study. The certificate exhibits both behaviours
a trustworthy go/no-go needs: it refuses the real target, which is classically trivial, and it
certifies the dense-spectrum substrate ($t{=}3$ headline; $t{=}2$ also certifies), where the
complete classical bar (LogReg $0.545$, GAM $0.548$, GA\textsuperscript{2}M $0.582$, HGB
$\doorCls$, JOINT $\doorJoint$; spread RFF at chance in screening) is beaten by the
dynamics-matched quantum model (DMQ) at identical information, $\Delta = \doorGap$
[CI $\doorGapCI$, $+0.310$]. The exact-simulation twin DMC ties DMQ ($\dmcAUC$) and is addressed
by the resource analysis of Figure~\ref{fig:projection}. The controlled sparse pocket of
Section~\ref{sec:control} is a bar-calibration instrument, not a certification candidate: its
known single term belongs to the order-matched twin ($\vOneJoint$ vs.\ $\vOneQNN$).}
\label{tab:verdicts}
\begin{tabular}{@{}llllll@{}}
\toprule
view & quantum & best classical & JOINT & gap & verdict \\
\midrule
real peak-regime target & \rawQNN & \textbf{\rawCls} (HGB) & -- & negative & \textsc{refuse} \\
dense-spectrum substrate (C1--C5) & \textbf{\doorQNN} & \doorCls\ (HGB) & \doorJoint & $\doorGap$ & \textsc{certify} \\
\bottomrule
\end{tabular}
\end{table*}

On the dense-spectrum substrate the quantum model beats \emph{every} tuned classical twin,
including the one built to close the exact gap the recipe targets; on the real target the ordering
flips to trees. Both behaviours are required: a detector that cannot refuse is marketing, and one
that cannot certify is pessimism.

\begin{figure}[tb]\centering
\includegraphics[width=\columnwidth]{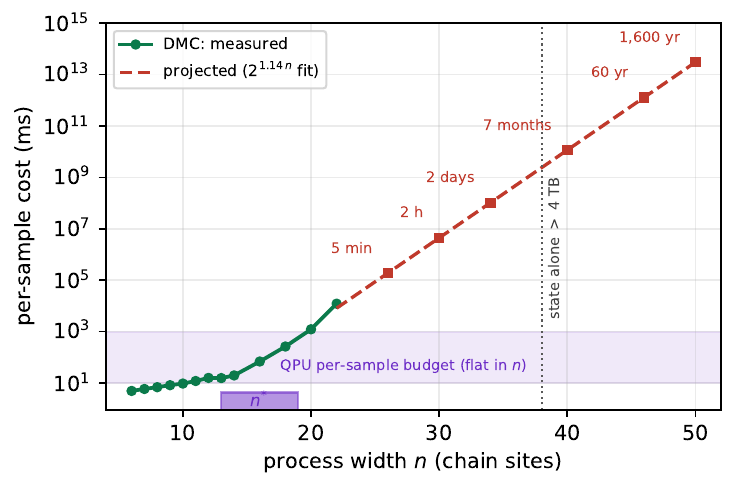}
\caption{What the accuracy tie costs, and what happens with more qubits. Measured per-sample cost
of exact classical simulation (green; the only classical evaluation route for this family) grows as
$2^{\simExp\, n}$ on the fitted tail (dashed): minutes at $n{=}26$, hours at $n{=}30$, months at
$n{=}40$, centuries at $n{=}50$, with the statevector alone exceeding $4$\,TB near $n{=}38$
(dotted). A quantum processor evaluates the same family at a near-constant per-sample budget
(shaded band: $10$\,ms--$1$\,s for $10^3$--$10^5$ shots at published gate
times~\cite{kim2023utility}); the curves cross at $n^{*}\approx 13$--$19$ sites (violet marker).
DMC ties DMQ's accuracy everywhere it can run; the plot is the price of continuing to run it.}
\label{fig:projection}
\end{figure}

\section{Mechanism evidence and deployment levers}
\label{sec:mechanism}

\paragraph{Entanglement is load-bearing: the certified substrate.} On the dense-spectrum substrate
the ablation is as clean as an ablation can be: clamping the interaction couplings to zero
($J{=}0$, product dynamics, the additive-model equivalent) collapses DMQ to chance. Under the
paired protocol (both arms retrained from scratch at identical budget, differences bootstrapped on
the same test rows) the ON$-$OFF gap is $\doorEntGap$ [CI $+0.469$, $+0.499$], with the OFF arm
statistically indistinguishable from $0.5$. Nothing else changes between the two configurations:
same encoding of the same real phases, same observable, same optimiser, same budget. This is the Fourier picture's prediction
run to its extreme: without the couplings the model's frequency support factorises
(Eq.~\ref{eq:additive}) and a dense joint spectrum is simply out of
reach~\cite{schuld2021effect,barthe2024reuploading}.

\paragraph{Entanglement in the encoding: the sparse-pocket measurement.} The sparse pocket's
ablation is informative even though the pocket itself belongs to the order-matched classical:
comparing two \emph{quantum} configurations is a statement about the mechanism, not about the
classical bar.
There, ablating the entangling \emph{encoding} (per-qubit encoding, all else fixed) cost $\entGap$
(CI $[+0.126,\,+0.147]$) while ablating the \emph{variational} entangler was null ($-0.004$,
CI $[-0.013,\,+0.005]$): entanglement pays where it changes the frequency support $\Omega$, not
where it remixes coefficients~\cite{sim2019expressibility}.

\begin{figure}[tb]\centering
\includegraphics[width=\columnwidth]{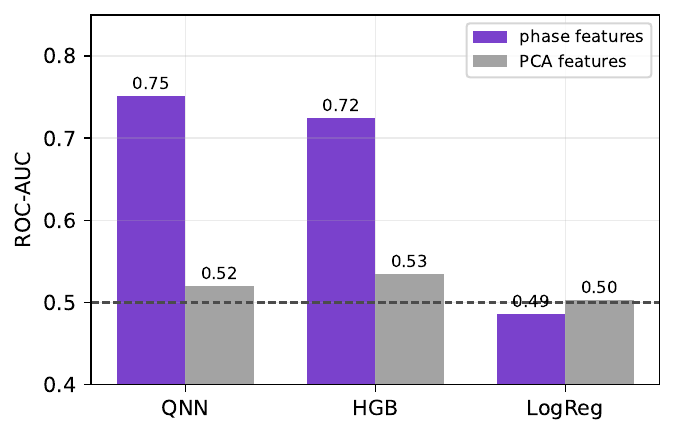}
\caption{The same three model families fed with periodic phase features (violet) versus
variance-maximising PCA components of the full feature block (grey), on the engineered substrate;
dashed line is chance. Every PCA lane collapses: the advantage-bearing structure lives on phase axes
that variance does not see.}
\label{fig:pca}
\end{figure}

\paragraph{Periodic phases beat PCA (measured on the sparse pocket).} Feeding the same
downstream models with variance-maximising PCA components of the full feature block, instead of the
periodic phases, collapses every lane to near-chance (Figure~\ref{fig:pca}): QNN $0.520$, HGB
$0.534$, logistic regression $0.503$. On phase features the same models reach QNN $0.752$ and HGB
$0.724$. This is a statement about \emph{representations}, not about the classical bar. The practical lesson for any industry pipeline: \emph{cyclic
covariates should be encoded as phases, not compressed by variance}. PCA is where periodic structure
goes to die.

\paragraph{No single-term niche exists at any frequency.} A sweep over controlled synthetic
single-term order-3 targets confirms the sparse-pocket lesson across the whole frequency axis: trees
collapse as the non-integer frequency rises (C4) and the bounded-order Fourier classicals plateau
(C2), yet the order-matched twin tracks the ceiling at every scale tested (ROC-AUC $0.97$--$0.99$ at
scales $1$ to $4.5$, against quantum-lane values of $0.61$--$0.65$ at matched budget). High frequency
defeats trees and high order defeats GA\textsuperscript{2}M, but as long as the spectrum is sparse,
enumeration wins. Every classical model class in the bar has a failure axis; only their intersection
(C1--C5 jointly) is quantum territory.

\begin{figure}[tb]\centering
\includegraphics[width=0.92\columnwidth]{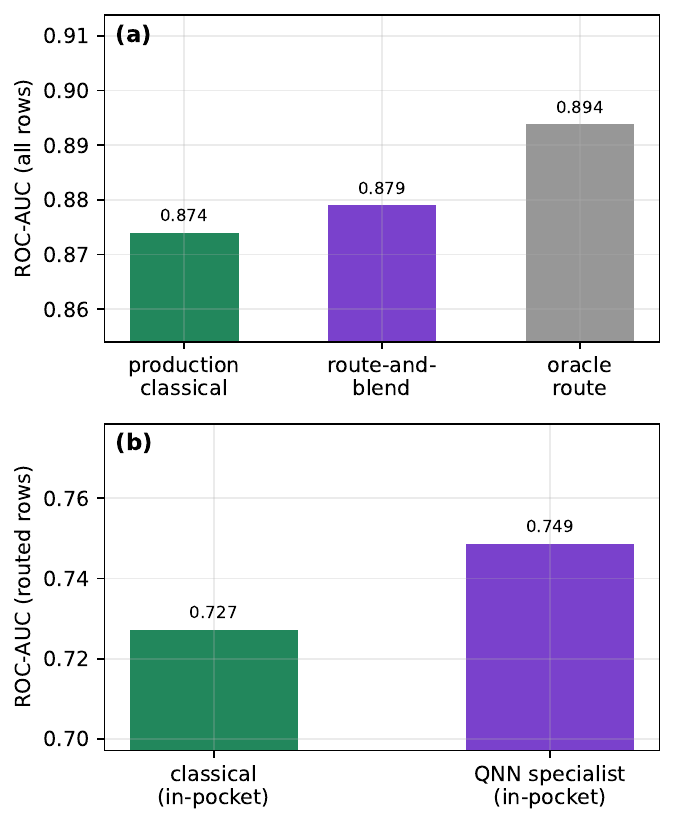}
\caption{Route-and-blend deployment on the energy substrate. \textbf{(a)} Global scores: the
production classical, the route-and-blend deployment, and the oracle router upper bound.
\textbf{(b)} On routed rows only: the certified QNN specialist against the production classical
inside the pocket. The router finds the pocket essentially perfectly (ROC-AUC $\approx 1.00$).}
\label{fig:routing}
\end{figure}

\paragraph{Routing: the certified specialist, not a replacement (architecture validated on the
sparse pocket).} No one should discard a production classifier to adopt a quantum model. The
deployment pattern is \emph{route-and-blend} (Algorithm~\ref{alg:route}): a cheap router identifies
the pocket of rows carrying certified structure, the certified specialist scores only those, and its
ordering is blended back into the production model's risk bands (Figure~\ref{fig:routing}). The
architecture was validated on the sparse pocket: the specialist won its rows ($0.749$ vs.\ $0.727$)
and the blend added $\routeBenefit$ [CI $\routeCIlo$, $\routeCIhi$] over the production classical
without touching any score outside the pocket. The architecture is model-agnostic: the specialist
slot takes \emph{whichever model the certificate certifies for the pocket}. On a sparse pocket that
model is JOINT; on a dense-spectrum pocket it is the dynamics-matched quantum model. Two requirements are
load-bearing regardless of the occupant. First, the router's gate must be \emph{independent} of the
specialist's features (gating on them re-correlates the pocket, violating C3, and the benefit
disappears). Second, the deployed specialist must match the certified configuration exactly; an
under-trained single-restart variant turned the same blend negative ($-0.043$) even with the pocket
advantage present.

\begin{algorithm}[t]
\caption{Route-and-blend deployment of a certified QNN specialist}
\label{alg:route}
\begin{algorithmic}[1]
\small
\State \textbf{given:} production classical $C$; router $R$
\Statex \hspace{\algorithmicindent} (gate independent of the QNN phases);
\Statex \hspace{\algorithmicindent} specialist $Q$ (certified config, in-pocket); band $b$
\For{each scoring row $x$}
  \If{$R(x) > \tau$} \Comment{row carries recipe structure}
     \State $s(x) \gets$ re-rank within $C$-band of width $b$ by $Q(x)$
  \Else
     \State $s(x) \gets C(x)$ \Comment{production score untouched}
  \EndIf
\EndFor
\State \Return $s$ \Comment{global calibration preserved}
\end{algorithmic}
\end{algorithm}

\section{Beyond energy: reading other industries through the recipe}
\label{sec:beyond}

The case study is energy, but the recipe is domain-blind, and the wall explains in advance what
practitioners across classification domains should expect from their data as exported.

\begin{itemize}
  \item \emph{Healthcare.} Vital signs and admissions data are rich in physiological and
        institutional cycles (circadian, weekly staffing, seasonal), but tabular clinical targets are
        typically driven by \emph{levels and trends}, i.e.\ low-order, smooth structure on strongly
        correlated covariates. C2 and C3 fail, and trees win. A door could open where several
        \emph{independent} biological rhythms interact jointly (chronotherapy response timing is a
        natural candidate), if encoded as phases rather than aggregates.
  \item \emph{Oil \& gas and process industry.} Rotating equipment, duty cycles, and maintenance
        calendars are genuinely multi-periodic, and failures often arise from \emph{coincidences} of
        cycles, a physically plausible source of C2-type joint structure. The recipe says exactly
        what the feature pipeline must preserve for that to become a QNN pocket: phase encodings per
        cycle and decorrelation of the encoded block (C3), rather than RMS or summary statistics that
        erase C1 and C4.
  \item \emph{Manufacturing quality and telecom operations.} Shift patterns, tool-wear cycles,
        traffic periodicities: the raw physics is cyclic, the standard exports are not. Aggregation
        to daily KPIs collapses the spectrum onto the grid and below tree frequency. Here the wall is
        an \emph{artifact of preprocessing}, which is good news: it is reversible.
\end{itemize}

The general translation: (1) re-express every cyclic mechanism as a phase; (2)
whiten the encoded block; (3) hunt for targets driven by joint timing of $\geq 3$
mechanisms rather than by levels; (4) run tier~1 of the certificate, minutes on a laptop,
before any quantum investment; (5) if a pocket certifies, deploy as a routed specialist
(Algorithm~\ref{alg:route}), never as a replacement.

\section{Honesty and scope}
\label{sec:honesty}

\begin{itemize}
  \item \emph{The bar is complete, and completeness is demonstrated, not asserted.} On the
        controlled sparse pocket, removing the order-matched twin from the lineup would credit
        the quantum model with a spurious advantage ($\vOneQNN$ vs.\ $\vOneThree$); with the twin
        present the structure is read classically outright ($\vOneJoint$), exactly as the recipe
        assigns it. Any advantage claim whose comparison set lacks an order-matched model should
        be read as untested.
  \item \emph{The certified advantage is on a semi-synthetic substrate and is labelled as such}:
        real energy covariate marginals, label generated by a quantum process on those covariates
        (the positive-control construction of~\cite{huang2021power}). It is a statement about
        \emph{which data} rewards a quantum model, not a claim that today's energy exports do. The
        real peak-regime target is refused, and that refusal is a result.
  \item \emph{Matched lanes on both sides, including the one that ties.} The classical twins
        get supervised structure search; DMQ gets the physics-informed family with learned
        parameters; and the strongest classical competitor of all, DMC, gets the \emph{identical}
        family and fit. DMC ties the accuracy and is reported, not hidden; what separates the lanes
        is the measured, exponentially-scaling cost of the classical evaluation
        (Figure~\ref{fig:projection}). Advantage claims where only one side is allowed an inductive
        bias are not claims~\cite{kubler2021inductive,bowles2024better}; neither are claims that
        omit the competitor which ties.
  \item \emph{Bounded, not provable, and with the boundary now mapped.} At $\leq 14$ qubits
        this is a certification methodology plus a measured resource-scaling argument, not an
        unconditional separation (which requires special structure~\cite{liu2021rigorous} or the
        tens-of-qubits regime~\cite{dynamics2025advantage}); at seven sites specifically, no
        hardness applies and DMC ties. The door closes on both sides: toward weak scrambling the
        classical bar climbs, toward strong scrambling everything becomes untrainable at finite
        budget (Figure~\ref{fig:regime}).
  \item \emph{Correlation dequantizes; enumeration dequantizes.} C3 is fragile in the wild, and
        C5 is the harder condition: any pocket whose structure a practitioner can \emph{write
        down} as a few terms is a pocket a classical model can fit. The advantage lives only where
        the spectrum is dense beyond enumeration.
  \item \emph{What survives scrutiny}: the Fourier taxonomy, the five necessary conditions, the
        two-tier certificate, the entanglement mechanism (a $\doorEntGap$
        ablation gap on the certified substrate), and the deployment pattern. All are reusable
        verbatim on any tabular classification dataset.
\end{itemize}

\section{Conclusion}
\label{sec:conclusion}

This study set out to explain a decade of null results on public tabular benchmarks and ends with
a certified, priced statement of where quantum advantage lives. Four findings carry the paper:

\begin{itemize}
  \item \emph{The wall is structural, and naming it is progress.} Some tuned classical model
        reaches every target whose spectrum is low-order, on-grid, slow, correlated, or sparse
        enough to enumerate, and public tabular exports essentially always land inside that reach.
        The negative benchmark literature is therefore evidence about \emph{datasets}, not about
        quantum models. Those contests were unwinnable by construction, and a five-condition test
        now says so in minutes, before any quantum cycle is spent.
  \item \emph{Advantage claims need a complete classical bar.} The order-matched twin, a dozen
        parameters fit in seconds, reads outright a controlled substrate that any bar lacking it
        would have credited to the quantum model ($\vOneJoint$ vs.\ $\vOneQNN$). Enumeration is
        the failure mode most published claims never test.
  \item \emph{A door exists, on real covariates.} Where the label is the trace of a genuinely
        interacting process, the dynamics-matched quantum model reaches $\doorQNN$ against the
        complete bar's $\doorCls$ ($\Delta = \doorGap$, CI-lo $\doorGapCI$) and collapses to
        chance the moment its couplings are ablated ($\doorEntGap$). Entanglement is the
        mechanism, not a decoration.
  \item \emph{The certified benefit is a measured resource separation.} The exact classical
        simulator of the family ties the accuracy at seven sites, as it must, and pays for the
        tie with a per-sample cost measured to grow as $2^{\simExp\,n}$ with process width, against a
        near-constant quantum-hardware budget, with the wall-clock crossover at
        $n^{*}\approx 13$--$19$ sites, just above the width of this study.
\end{itemize}

\paragraph{Implications for the next studies.} The findings convert directly into methodology.
Dataset choice should be spectral rather than availability-driven; tier-1 screening replaces the habit of
benchmarking on whatever export is public, and a refusal is a publishable calibration point rather
than a failed experiment. Every advantage claim should ship the complete bar, including the
order-matched twin and, whenever the generating family is known, the exact-simulator twin that
ties, with paired confidence bounds on held-out data. And the comparison should be fixed
\emph{before} the quantum model is trained, so that the verdict is pre-registered rather than
post-hoc. Under those rules, the productive research direction inverts, from tuning models against
data that cannot reward them to finding or engineering substrates (process telemetry of
interacting systems, sensing pipelines, deliberately recipe-conformant feature stacks) whose
spectra are dense beyond enumeration.

\paragraph{Implications for the quantum ecosystem.} For near-term hardware, the result relocates
the value proposition for machine learning. The offer is not higher accuracy at widths a laptop can simulate,
where the simulator twin always ties, but \emph{affordable evaluation} of dense joint spectra past
the classical crossover, where the tie becomes unpayable. That gives vendors a concrete workload
target (interacting-dynamics learning tasks at $n \gtrsim 20$ sites), and it gives adopters a
procurement instrument, a calibrated go/no-go on their own data before any hardware spend, which
protects the field's credibility as much as any budget. The certificate is the missing interface
between the two.

\paragraph{Outlook: hardware validation.} The immediate next step is already defined by the
measurements above, and we intend to take it as soon as hardware access allows. The run is DMQ on a
quantum processor for the same disordered-chain learning task at $n \approx 20$--$40$ sites, the
window just past the measured crossover $n^{*}$, where the exact classical twin is no longer
affordable and the certified accuracy separation over every remaining classical model becomes an
empirical fact rather than a priced projection. Figure~\ref{fig:projection} extrapolates the
measured wall to make the window concrete. On the fitted tail, one classical evaluation costs about
five minutes at $n{=}26$, two hours at $n{=}30$, seven months at $n{=}40$, and sixteen centuries at
$n{=}50$, with the statevector alone outgrowing terabytes near $n{=}38$, while the quantum budget
stays flat. The required circuit widths and two-qubit-gate depths sit inside envelopes already
demonstrated with error mitigation on current devices~\cite{kim2023utility}, and the certificate
developed here is precisely the instrument that makes such a run decisive, since tier~1 selects the
dataset, tier~2 fixes the complete classical bar in advance, and the verdict is a pre-registered,
paired comparison rather than a post-hoc benchmark.

\paragraph{Implementation availability.} The SPECTRA certificate, the substrate machinery, and the
routed-deployment stack are proprietary technology of Falcondale, currently under active product
development. Organisations interested in screening their datasets, engineering recipe-conformant
features, or piloting a routed quantum specialist can contact us at
\href{mailto:contact@falcondale.pro}{\texttt{contact@falcondale.pro}}.

\section*{Acknowledgments}

We are deeply grateful to Iraitz Montalb\'an, whose contribution to this study was
significant and far-reaching: he shaped the analysis, the design and completion of the classical
benchmark bar, the theoretical framing of the Fourier-wall argument, and the reading of its
implications for practitioners and for the quantum ecosystem. This paper is substantially stronger
for his insight and rigour at every one of those levels. We also warmly thank
Christophe Pere for his generous and incisive peer review; his careful reading and
demanding standards directly improved the completeness of the classical comparison and the clarity
of the claims. Any remaining errors are, of course, our own.

\clearpage   
\bibliographystyle{unsrt}
\bibliography{references}

@article{schuld2021effect,
  title={The effect of data encoding on the expressive power of variational quantum-machine-learning models},
  author={Schuld, Maria and Sweke, Ryan and Meyer, Johannes Jakob},
  journal={Physical Review A}, volume={103}, number={3}, pages={032430}, year={2021},
  note={arXiv:2008.08605}}

@article{jaderberg2024trainable,
  title={Let quantum neural networks choose their own frequencies},
  author={Jaderberg, Ben and Gentile, Antonio A. and Achari Berrada, Youssef and
          Shishenina, Elvira and Elfving, Vincent E.},
  journal={Physical Review A}, volume={109}, pages={042421}, year={2024}, note={arXiv:2309.03279}}

@article{longrange2026,
  title={Long range frequency tuning for {QML}},
  author={Poppel, Michael and Baumann, Markus and W{\"o}lckert, Sebastian and
          Linnhoff-Popien, Claudia and Stein, Jonas},
  journal={arXiv preprint arXiv:2602.23409}, year={2026}}

@article{barthe2024reuploading,
  title={Gradients and frequency profiles of quantum re-uploading models},
  author={Barthe, Alice and P{\'e}rez-Salinas, Adri{\'a}n},
  journal={Quantum}, volume={8}, pages={1523}, year={2024}, note={arXiv:2311.10822}}

@article{bowles2024better,
  title={Better than classical? {T}he subtle art of benchmarking quantum machine learning models},
  author={Bowles, Joseph and Ahmed, Shahnawaz and Schuld, Maria},
  journal={arXiv preprint arXiv:2403.07059}, year={2024}}

@article{schuld2022advantage,
  title={Is quantum advantage the right goal for quantum machine learning?},
  author={Schuld, Maria and Killoran, Nathan},
  journal={PRX Quantum}, volume={3}, pages={030101}, year={2022}, note={arXiv:2203.01340}}

@inproceedings{kubler2021inductive,
  title={The inductive bias of quantum kernels},
  author={K{\"u}bler, Jonas M. and Buchholz, Simon and Sch{\"o}lkopf, Bernhard},
  booktitle={Advances in Neural Information Processing Systems (NeurIPS)}, year={2021},
  note={arXiv:2106.03747}}

@article{huang2021power,
  title={Power of data in quantum machine learning},
  author={Huang, Hsin-Yuan and Broughton, Michael and Mohseni, Masoud and Babbush, Ryan and
          Boixo, Sergio and Neven, Hartmut and McClean, Jarrod R.},
  journal={Nature Communications}, volume={12}, number={1}, pages={2631}, year={2021},
  note={arXiv:2011.01938}}

@article{sweke2023rff,
  title={Potential and limitations of random {F}ourier features for dequantizing quantum machine learning},
  author={Sweke, Ryan and Recio-Armengol, Erik and Jerbi, Sofiene and Gil-Fuster, Elies and
          Fuller, Bryce and Eisert, Jens and Meyer, Johannes Jakob},
  journal={arXiv preprint arXiv:2309.11647}, year={2023}}

@inproceedings{gilfuster2024relation,
  title={On the relation between trainability and dequantization of variational quantum learning models},
  author={Gil-Fuster, Elies and Gyurik, Casper and P{\'e}rez-Salinas, Adri{\'a}n and Dunjko, Vedran},
  booktitle={International Conference on Learning Representations (ICLR)}, year={2025},
  note={arXiv:2406.07072}}

@article{hastie1986gam,
  title={Generalized additive models},
  author={Hastie, Trevor and Tibshirani, Robert},
  journal={Statistical Science}, volume={1}, number={3}, pages={297--310}, year={1986}}

@inproceedings{lou2013ga2m,
  title={Accurate intelligible models with pairwise interactions},
  author={Lou, Yin and Caruana, Rich and Gehrke, Johannes and Hooker, Giles},
  booktitle={Proceedings of the 19th ACM SIGKDD International Conference on Knowledge Discovery and
             Data Mining (KDD)}, pages={623--631}, year={2013}}

@article{friedman2001gbm,
  title={Greedy function approximation: A gradient boosting machine},
  author={Friedman, Jerome H.},
  journal={The Annals of Statistics}, volume={29}, number={5}, pages={1189--1232}, year={2001}}

@inproceedings{ke2017lightgbm,
  title={{LightGBM}: A highly efficient gradient boosting decision tree},
  author={Ke, Guolin and Meng, Qi and Finley, Thomas and Wang, Taifeng and Chen, Wei and Ma, Weidong
          and Ye, Qiwei and Liu, Tie-Yan},
  booktitle={Advances in Neural Information Processing Systems (NeurIPS)}, year={2017}}

@inproceedings{chen2016xgboost,
  title={{XGBoost}: A scalable tree boosting system},
  author={Chen, Tianqi and Guestrin, Carlos},
  booktitle={Proceedings of the 22nd ACM SIGKDD International Conference on Knowledge Discovery and
             Data Mining (KDD)}, pages={785--794}, year={2016}, note={arXiv:1603.02754}}

@inproceedings{grinsztajn2022trees,
  title={Why do tree-based models still outperform deep learning on typical tabular data?},
  author={Grinsztajn, L{\'e}o and Oyallon, Edouard and Varoquaux, Ga{\"e}l},
  booktitle={Advances in Neural Information Processing Systems (NeurIPS), Datasets and Benchmarks},
  year={2022}, note={arXiv:2207.08815}}

@inproceedings{rahimi2007rff,
  title={Random features for large-scale kernel machines},
  author={Rahimi, Ali and Recht, Benjamin},
  booktitle={Advances in Neural Information Processing Systems (NeurIPS)}, year={2007}}

@article{sim2019expressibility,
  title={Expressibility and entangling capability of parameterized quantum circuits for hybrid
         quantum-classical algorithms},
  author={Sim, Sukin and Johnson, Peter D. and Aspuru-Guzik, Al{\'a}n},
  journal={Advanced Quantum Technologies}, volume={2}, number={12}, pages={1900070}, year={2019},
  note={arXiv:1905.10876}}

@article{spectralamp2024,
  title={The spectral amplitude principle for dynamics of quantum neural networks},
  author={Xu, Yi-hang and Zhang, Dan-Bo and Yan, Junchi},
  journal={arXiv preprint arXiv:2409.06682}, year={2024}}

@article{liu2021rigorous,
  title={A rigorous and robust quantum speed-up in supervised machine learning},
  author={Liu, Yunchao and Arunachalam, Srinivasan and Temme, Kristan},
  journal={Nature Physics}, volume={17}, number={9}, pages={1013--1017}, year={2021}}

@article{dynamics2025advantage,
  title={Evidence of quantum machine learning advantage with tens of noisy qubits},
  author={Danaci, Onur and Patel, Yash J. and Molteni, Riccardo and van Nieuwenburg, Evert and
          Dunjko, Vedran and Krzywda, Jan A.},
  journal={arXiv preprint arXiv:2605.21346}, year={2026}}

@misc{uci_steel,
  title={Steel Industry Energy Consumption},
  author={Sathishkumar, V. E. and Shin, Changsun and Cho, Yongyun},
  howpublished={UCI Machine Learning Repository, id 851}, year={2021},
  note={DOI: 10.24432/C52G8C. \url{https://archive.ics.uci.edu/dataset/851}}}

@article{sathishkumar2020building,
  title={Efficient energy consumption prediction model for a data analytic-enabled industry
         building in a smart city},
  author={Sathishkumar, V. E. and Shin, Changsun and Cho, Yongyun},
  journal={Building Research \& Information}, volume={49}, number={1}, pages={127--143},
  year={2021}, note={DOI: 10.1080/09613218.2020.1809983}}

@article{sathishkumar2020steel,
  title={An energy consumption prediction model for smart factory using data mining algorithms},
  author={Sathishkumar, V. E. and Lee, Myeongbae and Lim, Jonghyun and Kim, Yubin and
          Shin, Changsun and Park, Jangwoo and Cho, Yongyun},
  journal={KIPS Transactions on Software and Data Engineering}, volume={9}, number={5},
  pages={153--160}, year={2020}, note={DOI: 10.3745/KTSDE.2020.9.5.153}}

@article{kim2023utility,
  title={Evidence for the utility of quantum computing before fault tolerance},
  author={Kim, Youngseok and Eddins, Andrew and Anand, Sajant and Wei, Ken Xuan and
          van den Berg, Ewout and Rosenblatt, Sami and Nayfeh, Hasan and Wu, Yantao and
          Zaletel, Michael and Temme, Kristan and Kandala, Abhinav},
  journal={Nature}, volume={618}, pages={500--505}, year={2023}}

\end{document}